\documentclass[aps,prd,twocolumn,10pt,superscriptaddress,showpacs,floatfix,nofootinbib]{revtex4-1}
\usepackage{amssymb}
\usepackage{textcomp}
\usepackage{graphicx}
\usepackage{epsfig}
\usepackage{amsmath}
\usepackage{slashed}
\usepackage{units}
\usepackage{hyperref}
\usepackage{array}
\usepackage{lmodern}

\begin{document}

\title{Probing the braneworld hypothesis with a neutron-shining-through-a-wall experiment}

\author{Micha\"{e}l Sarrazin}
\email{michael.sarrazin@unamur.be} \affiliation{Solid-State Physics Laboratory, Research Center in Physics of Matter and Radiation, 
University of Namur, 61 rue de Bruxelles, B-5000 Namur, Belgium}

\author{Guillaume Pignol}
\affiliation{LPSC, Universit\'{e} Grenoble-Alpes, CNRS/IN2P3, 53 rue des Martyrs, F-38026 Grenoble, France}

\author{Jacob Lamblin}
\affiliation{LPSC, Universit\'{e} Grenoble-Alpes, CNRS/IN2P3, 53 rue des Martyrs, F-38026 Grenoble, France}

\author{Fabrice Petit}
\affiliation{BCRC (Member of EMRA), 4 avenue du Gouverneur Cornez, B-7000 Mons, Belgium}

\author{Guy Terwagne}
\affiliation{Laboratory of Analysis by Nuclear Reactions, Research Center in Physics of Matter and Radiation,
University of Namur, 61 rue de Bruxelles, B-5000 Namur, Belgium}

\author{Valery V. Nesvizhevsky}
\affiliation{Institut Laue-Langevin, 6 rue Jules Horowitz, F-38042 Grenoble,
France}

\begin{abstract}
The possibility for our visible world to be a 3-brane embedded in a
multidimensional bulk is at the heart of many theoretical edifices in high-energy 
physics. Probing the braneworld hypothesis is thus a major
experimental challenge. Following recent theoretical works showing that
matter swapping between braneworlds can occur, we propose a \emph{%
neutron-shining-through-a-wall} experiment. We first show that an intense
neutron source such as a nuclear reactor core can induce a hidden neutron
flux in an adjacent hidden braneworld. We then describe how a low-background
detector can detect neutrons arising from the hidden world and quantify the
expected sensitivity to the swapping probability. As a proof of concept, a
constraint is derived from previous experiments.
\end{abstract}


\maketitle

\section{Introduction}

As suggested by several extensions of the standard models of particle physics
and cosmology, our world could form a three-dimensional space sheet (a
brane) embedded in a larger bulk with more dimensions \cite{1,2,3}. As for
most theories beyond the Standard Model of particle physics, new effects
predicted by the braneworld hypothesis at both the high-energy and the
precision frontiers can be investigated. Particle colliders, such as the
Large Hadron Collider, attempt to excite new degrees of freedom at high
energies \cite{4,5} -- like the Kaluza-Klein excitations of particles in the
extra dimensions -- whereas precision experiments at low energies attempt to
find tiny signals induced by the new physics \cite
{6,7,7b,8a,8b,8,11,12,11b,13,14,15}. For instance, in the context of braneworld
scenarios, the compactified extra dimensions would modify the inverse-square
law of gravity at short distance. A great variety of experimental techniques
have been developed to search for a possible modification of gravity from
subatomic to macroscopic distance scales (see Ref. \cite{6} for a recent review).

In the present work, we are interested in some peculiar low-energy effects
induced by the existence of other branes in the bulk. In recent theoretical
works \cite{11,12,11b,13,14,15}, it has been argued that usual matter could
leap from our braneworld to a hidden one, and vice versa (see Fig. 1). This
matter swapping between two neighboring braneworlds could be triggered by
magnetic vector potentials, either of astrophysical origin or generated
experimentally. In particular, we have considered neutrons oscillating
between a state where they sit in our brane and a state corresponding to a
neutron located in the other brane. The matter-swapping probability would
oscillate at high frequency $\eta$ and small amplitude $p$. Moreover,
neutrons are well-known versatile tools which are already used to test other
concepts such as axion \cite{15b} or mirror particles \cite{15c,15d,15e,15ee,15f}.

In a previous work, we have shown that such oscillations could have affected
experiments measuring the neutron lifetime. The existing experiments set a
constraint on anomalous neutron \emph{disappearance} at the level of $%
p<7\times 10^{-6}$ \cite{15}. It is amusing to note that a small tension has
very recently appeared between the neutron lifetimes measured with two
different methods \cite{9}, which could be interpreted as anomalous neutron
disappearance. In addition, such a constraint on neutron disappearance could
also be useful to constrain some approaches related to the big bang
nucleosynthesis \cite{10}.

In the present paper, we propose an experiment to investigate matter
swapping between branes by looking at the \emph{appearance} of neutrons from
a neighboring brane. The concept is similar to \emph{%
light-shining-through-a-wall} experiments (see Ref. \cite{7} for a recent review
of this topic) where photons from an intense light source would convert into
a sterile state (dark photons or axion-like particles) which could pass
through a wall, then would convert back to photons. In a \emph{%
neutron-shining-through-a-wall} experiment, a very bright source of neutrons
and a low-background neutron detector separated by a wall are necessary.
Neutrons would swap into a sterile state -- the state where they are located
in another brane -- which would be free to cross the wall. Then, the
reappearance of neutrons into the detector situated behind the wall is
checked. Although they are in principle very similar, there is an
essential difference between light and neutrons shining through walls. As
far as light is concerned, the oscillation frequency is supposed to be
rather slow and the conversion of photons into a sterile state builds up
coherently over long distance, while the oscillation is rather fast for
neutrons and their conversion into sterile states results from the
successive collisions of neutrons at nuclei.

In section II, the theoretical aspects and the phenomenology of the neutron
dynamics in a two-brane universe at low energy are reviewed. The conditions
leading to matter swapping between branes are given. The strength of the
ambient magnetic vector potential (which drives the matter exchange between
branes) is discussed, as well as the environmental conditions that could
preclude the swapping to occur. In section III, we describe how the neutron
diffusion in the reactor moderator creates a neutron flux in another brane
and how this hidden neutron flux can be detected. Finally, the sensitivity
of the suggested experiment is estimated, and as a proof of the concept, a
constraint is derived from previous experiments.

\section{Phenomenology of brane matter swapping}

\label{model}

\subsection{Dynamics of swapping}

\begin{figure}[th]
\centerline{\ \includegraphics[width=4.5cm]{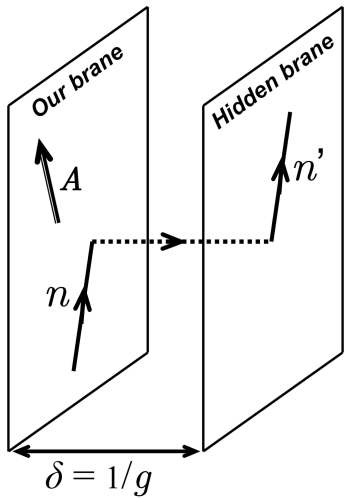}}
\caption{Naive view of the neutron swapping between two branes. A neutron $n$
in our brane can be transferred into a hidden brane (neutron $n^{\prime }$)
under the influence of a suitable magnetic vector potential $\textbf{A}$%
. $g$ represents the coupling strength between the branes related to the
effective distance $\delta$ between each brane.}
\label{fig1}
\end{figure}

In previous works \cite{11,12,11b,13}, it was shown that, in a universe
containing two parallel braneworlds, the quantum dynamics of a spin$-1/2$
fermion in the presence of an electromagnetic field can be described by a
two-brane Pauli equation at low energies. In electromagnetic and
gravitational fields, the dynamics of a free neutron, i.e., outside of a
nucleus, is described by the following equation \cite{11,12,11b}: 
\begin{equation}
i\hbar \frac \partial {\partial t}\left( 
\begin{array}{c}
\psi _{+} \\ 
\psi _{-}
\end{array}
\right) =\left( \mathbf{H}_0+\mathbf{H}_{cm}\right) \left( 
\begin{array}{c}
\psi _{+} \\ 
\psi _{-}
\end{array}
\right),  \label{2BPauli}
\end{equation}
where the indices $\pm $ allow one to discriminate the two branes. $\psi _{+}$
and $\psi _{-}$ are usual Pauli spinors corresponding to the wave functions
in the $(+)$ and $(-)$ branes, respectively. The unperturbed Hamiltonian $%
\mathbf{H}_0$ is block diagonal and describes the usual nonrelativistic evolution
of two uncoupled spin-1/2 fields in the magnetic fields $\mathbf{B}_{\pm
} $ and the gravitational potentials $V_{\pm }$, in each brane:

\begin{equation}
\mathbf{H}_0=\left( 
\begin{array}{cc}
\mathbf{H}_{+} & 0 \\ 
0 & \mathbf{H}_{-}
\end{array}
\right) ,  \label{UsualPauli}
\end{equation}
\begin{equation}
\mathbf{H}_{\pm }=-\frac{\hbar ^2}{2m}\Delta -\mu _n\mathbf{\sigma \cdot B}%
_{\pm }+V_{\pm }.  \label{Pau}
\end{equation}
The nondiagonal part $\mathbf{H}_{cm}$ of the Hamiltonian describes the
coupling between the branes, 
\begin{equation}
\mathbf{H}_{cm}=g\mu _n\left( 
\begin{array}{cc}
0 & -i\mathbf{\sigma \cdot }\left\{ \mathbf{A}_{+}-\mathbf{A}_{-}\right\} 
\\ 
i\mathbf{\sigma \cdot }\left\{ \mathbf{A}_{+}-\mathbf{A}_{-}\right\}  & 0
\end{array}
\right) ,  \label{Coupling}
\end{equation}
where $\mathbf{A}_{+}$ and $\mathbf{A}_{-}$ correspond to the magnetic
vector potentials in the branes $(+)$ and $(-)$, respectively. The same
convention is applied to the magnetic fields $\mathbf{B}_{\pm }$ and to the
gravitational potentials $V_{\pm }$. $\mu _n$ is the magnetic moment of the
neutron. $\mathbf{H}_{cm}$ implies that matter exchange between branes
depends on the magnetic moment and on the difference between the local (i.e.,
on a brane) values of the magnetic vector potentials. $g$ is the coupling
strength between the matter fields of each brane.

In the following section, the case of an ambient magnetic potential with an
astrophysical origin is considered (section \ref{Magvect}). Let $\mathbf{A}%
_{amb}=\mathbf{A}_{amb,+}-\mathbf{A}_{amb,-}$ be the difference between the ambient
magnetic potentials of each brane. Assuming that $\mu _nB_{\pm }\ll V_{\pm }$%
, i.e.,  the magnetic fields in the branes can be neglected (in particular
assuming that $\mathbf{\nabla }\times \mathbf{A}_{amb}=\mathbf{0}$), then by
solving the Pauli equation, the probability for a neutron initially
localized in our brane to be found in the other brane is \cite{11} 
\begin{equation}
P=\frac{4\Omega ^2}{\eta ^2+4\Omega ^2}\sin ^2\left( (1/2)\sqrt{\eta
^2+4\Omega ^2}t\right),  \label{Rabi}
\end{equation}
where $\eta =|V_{+}-V_{-}|/\hbar $ and $\Omega =g\mu _nA_{amb}/\hbar $. $P$
is the instantaneous probability for matter swapping between branes. Equation (%
\ref{Rabi}) shows that the neutron in the potential $A_{amb}$ undergoes
Rabi-like oscillations between the branes. Note that the swapping
probability is independent of the neutron spin direction \cite{11b}. As
detailed in previous papers \cite{11b,13,14}, the environmental interactions
(related to $V_{\pm }$) are usually strong enough and the oscillations are
suppressed by the factor $\Omega /\eta <<1$. As a consequence, in a nucleus,
a neutron is fully frozen in its brane due to the large nuclear potential.

\subsection{Ambient magnetic vector potential}

\label{Magvect}

The overall ambient astrophysical magnetic vector potential $\mathbf{A}_{amb}$
was previously assessed in the literature \cite{16}. $\mathbf{A}_{amb}$ is the
sum of all of the magnetic vector potential contributions related to the magnetic fields of
astrophysical objects (planets, stars, galaxies, etc.) since $\mathbf{B}(%
\mathbf{r})=\mathbf{\nabla }\times \mathbf{A}(\mathbf{r})$. As a rule of
thumb, $A\approx DB$ where $D$ is the distance from the astrophysical source
and $B$ is the typical field induced by the object. At large distances from
sources (for instance, close to the Earth), $\mathbf{A}_{amb}$ is almost uniform
(i.e. $\mathbf{\nabla }\times \mathbf{A}_{amb}\approx \mathbf{0}$) and
cannot be canceled with magnetic shields \cite{16}. Now, Eqs. (\ref{Coupling}%
) and (\ref{Rabi}) show the dependence of the swapping effect against $%
\mathbf{A}_{amb}=\mathbf{A}_{amb,+}-\mathbf{A}_{amb,-},$ i.e. the difference
between the vector potentials of the two braneworlds. Since $\mathbf{A}%
_{amb,-}$ depends on unknown sources in the hidden brane, we cannot assess
its value. Then, $\mathbf{A}_{amb}$ should be considered as an unknown
parameter of the model. Nevertheless, the expected order of magnitude of $%
A_{amb}$ can be roughly constrained by $A_{amb,+}$ in our visible world
(since $A_{amb}$ results from a vectorial difference, it seems quite
unlikely that $A_{amb}$ can fortuitously fall to zero). Galactic magnetic field 
variations on local scales towards the
Milky Way core ($A_{amb}\approx 2\times 10^9$ T m) are usually assumed \cite
{16}. By contrast, the Earth's magnetic field leads to $200$ T m while
the Sun contributes $10$ T m \cite{16}. By contrast, intergalactic
contributions were expected to be about $10^{12}$\ T m \cite{16} (see
also our previous papers for a more detailed discussion \cite{13,14,15}).
Anyway, for now $A_{amb}$ is a parameter fairly bounded between $10^9$ T m and $10^{12}$ T m.

\subsection{Environmental potential}

\label{envpot}

If one considers free neutrons shielded from magnetic fields, only
gravitational contributions are relevant. Because $\eta =|V_{+}-V_{-}|/\hbar 
$, it is difficult to assess the value of $\eta \hbar$ as it results from a
scalar difference involving the unknown gravitational contribution $V_{-}$
of the hidden world. Therefore, $\eta$ appears as an effective unknown
parameter of the model and could reach weak values of a few eV up to
large values around $1$ keV. Indeed, estimations given in previous works 
\cite{13,14,15} suggest that $V_{+}$ could be of the order of $500$ eV due
to the Milky Way core gravitational influence on neutrons. Note that the
Sun, the Earth, and the Moon provide lower contributions of about $9$ eV, $%
0.65$ eV, and $0.1$ meV, respectively. At last, it must be emphasized that $%
\eta$ is time dependent due the motion of the Earth around the Sun at the
lab scale. Between the Earth's aphelion and perihelion, the gravitational
energy (due to the Sun) of a neutron varies from $9.12$ to $9.43$ eV. Of
course, a time dependence could have many different origins. For instance,
the particle motion relative to an unknown mass distribution in the hidden
brane should be considered. However, it is unlikely that the Earth is
''close'' enough to a hidden mass distribution that is large enough
to induce a significant energy time dependence on a time scale of about one
year or one day. The time dependence is then mainly induced by the Earth's
motion around the Sun, with a relative variation in $\eta $ of about $0.3 \text{\textperthousand}$ 
in one year. Such a variation could be detected through an annual
modulation of the swapping probability (see section \ref{drift}).

\subsection{Neutrons as a sensitive probe}

With high-energy particle colliders, the braneworld hypothesis can be
investigated if the brane thickness is in the range $\xi \approx 10^{-19}$~m
corresponding to the TeV scale ($\hbar c/\xi \approx 1$~TeV). Colliders are
blind to the Planck scale $10^{-35}$~m. By contrast, experiments at lower
energies using high-intensity neutron sources could reveal a multibrane
world through effects induced by the interbrane coupling $g$, which can be
approximated by \cite{11}

\begin{equation}
g\propto (1/\xi )\exp \left( -kd/\xi \right),  \label{gcoup}
\end{equation}
where $d$ is the real distance between each brane in the bulk and $k$ is a
constant of the model \cite{11}. Now, as an illustration, let us consider (for instance)
 a coupling constant $g\approx $10$^{-3}$ m$^{-1}$. Such a
value is consistent with present experimental bounds \cite{15}. For branes
at the TeV scale, the above value of $g$ is reached for $d\approx 50\xi $.
Now, if one considers branes at the Planck scale, the coupling constant
remains unchanged for $d\approx 87\xi $. As a consequence, while brane
physics could be invisible for colliders, it could be observed in low-energy
experiments using neutrons if a second brane exists close enough to ours.

At last, we note that neutrons are more suitable than electrons, protons, or
atoms for such a purpose \cite{13}. Indeed, for a charged particle, the
Hamiltonian (\ref{Pau}) also contains the usual terms,

\begin{equation}
\mathbf{H}_{p,\pm }=-\frac qm\mathbf{A}_{\pm }\cdot \widehat{\mathbf{P}}+%
\frac{q^2}{2m}\left| \mathbf{A}_{\pm }\right| ^2 , \label{supp}
\end{equation}
with $\widehat{\mathbf{P}}=-i\hbar \mathbf{\nabla }$. This implies that the term 
$\hbar \eta $ in Eq. (\ref{Rabi}) is then supplemented by large terms
proportional to $qA_{amb}$ which strongly freeze the oscillations.
Considering, for instance, $A_{amb}\approx 200$ T m (see section \ref
{Magvect}) and a proton with a kinetic energy of about $10$ eV, the first
term in the Hamiltonian (\ref{supp}) contributes $9$ MeV to $\hbar \eta $.
For such values, the amplitude of the oscillations is suppressed by $8$ 
orders of magnitude compared to the neutron case. Now, let us consider
atoms. Though they are neutral, atoms are endowed with an instantaneous
electric dipole moment (IEDM) $\mathbf{d}$. Obviously, according to the time
average, we must verify that $\left\langle \mathbf{d}\right\rangle =\mathbf{0}$
and $\left\langle \mathbf{d}^2\right\rangle \neq \mathbf{0}$ since IEDM
results from quantum fluctuations of atomic orbitals. The London dispersion
forces between atoms result from interactions between these instantaneous
dipoles. Then, the Hamiltonian (\ref{Pau}) must be supplemented by terms
which derive from Eq. (\ref{supp}): 
\begin{equation}
\mathbf{H}_{d,\pm }\sim -\mathbf{A}_{\pm }\cdot \frac{\Delta \mathbf{d}}{%
\Delta t} , \label{supp2}
\end{equation}
where the fluctuation time is about $\Delta t\sim \hbar /E_i$, where $E_i$ is 
the ionization energy of the atom ($E_i\approx 10$ eV). Provided that $%
\Delta t$ is larger than the period $T$ of the Rabi oscillation of an atom
between two branes, $\hbar \eta $ in Eq. (\ref{Rabi}) is then supplemented
by a huge term proportional to $(1/\Delta t)\mathbf{A}_{amb}\cdot \mathbf{d}$%
. Then, for $d$ about $10^{-30}$ C$\cdot $m ($0.3$ D) (a typical value for
atoms) and $A_{amb}\approx 200$ T$\ $m only (for instance), we still get $%
\hbar \eta \approx 20$ MeV. With these values, the oscillations are strongly
damped by $9$ orders of magnitude compared to the neutron case. As a
result, the neutron is a good candidate to test matter swapping between
branes, since it is devoid of global charge or any electric dipole moment.

\subsection{Collision-induced neutron swapping probability}

\label{swp}

As a consequence of the environmental potential $\eta \hbar $, the neutron
oscillations present weak amplitude and high angular frequency of the order $%
\eta /2$. Due to the fast oscillating behavior, one can approximate the
swapping probability by its time-averaged value $p=\left\langle
P\right\rangle $ (see Eq. (\ref{Rabi})), such that
\begin{equation}
p=\frac{2\Omega ^2}{\eta ^2}.  \label{Proba}
\end{equation}
When freely propagating, the neutron can be described as a superposition of
two states: a neutron in our brane vs a neutron in the other brane. When
colliding with a nucleus situated in our brane, the interaction acts as a
measurement and the neutron collapses either in our brane with a probability 
$1-p$ or in the other invisible brane with a probability $p$. In the
following sections, the swapping probability $p$ is considered as the
relevant measurable parameter. By contrast, bounds on the coupling parameter 
$g$ depend on the knowledge of galactic magnetic potential fields and on the
ambient gravitational fields.

\section{A neutron-shining-through-a-wall experiment}

\begin{figure}[th]
\centerline{\ \includegraphics[width=8.5cm]{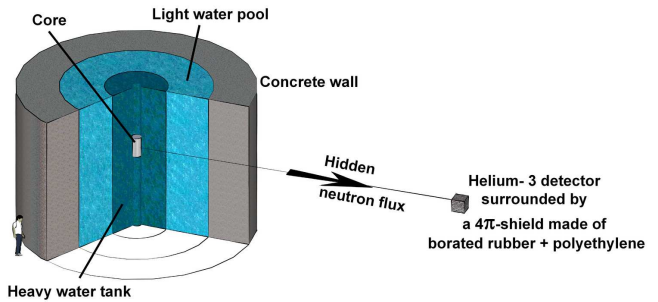}}
\caption{Sketch of the experimental device. The source of possible hidden
neutrons is a nuclear reactor core (here for instance, the Institut
Laue-Langevin facility in Grenoble, France). The neutron detector using
helium-3 gas is considered to detect neutrons which could emerge from a
hidden world. The detector is embedded in a shield to reach a very low
background.}
\label{fig2}
\end{figure}

As explained above, in a two-brane universe, neutrons have a nonzero
probability to escape from our brane into another brane at each collision.
Therefore, a nuclear reactor, where the neutron density is very high, would
be a very intense source of hidden neutrons. These neutrons could escape the
reactor and be detected, after having swapped back to our brane, with a
standard neutron detector located near the reactor (see Fig. 2). In the
present section, the expected magnitude of the hidden neutron flux as a
function of the swapping probability is discussed and the sensitivity which
can be reached with such an experiment is estimated. For sake of
clarity, the details of our calculations only appear in the Appendices.\\

\subsection{Induced neutron flux in the hidden world}

Let us assume that our Universe is made of two mutually invisible braneworlds and
let us also consider a neutron flux $\Phi _{+}$ inside a nuclear core. Our
aim is to determine the intensity of the hidden neutron flux $\Phi _{-}$ (in
the vacuum of the hidden braneworld) in the core vicinity. For a given
volume element of the reactor, since each neutron collision creates a
hidden neutron with probability $p$, the hidden neutron source is
proportional to the macroscopic elastic cross section $\Sigma _E$ of the
reactor moderator and to the neutron flux $\Phi _{+}$. More specifically, we
get the source term corresponding to the number of generated hidden neutrons
per unit volume and unit time (see Appendix \ref{Scatt}): 
\begin{equation}
S_{-}=\frac 12p\Sigma _E\Phi _{+},  \label{N1}
\end{equation}
where $p$ is given by Eq. (\ref{Proba}). Equation (\ref{N1}) is derived by using the
matrix density approach \cite{23,24} related to Eq. (\ref{2BPauli})
as shown in Appendix \ref{Scatt}. From this source term, we deduce the
hidden neutron flux $\Phi _{-}$ at the position $\mathbf{r}$ by considering
the solid angle and integrating over the reactor volume $V$: 
\begin{equation}
\Phi _{-}(\mathbf{r})=\frac p{8\pi }\int_V\frac 1{\left| \mathbf{r-r}%
^{\prime }\right| ^2}\Sigma _E(\mathbf{r}^{\prime })\Phi _{+}(\mathbf{r}%
^{\prime })d^3r^{\prime}.  \label{fluxn}
\end{equation}
The relation (\ref{fluxn}) shows that any signal should decay as $1/D^2$
when the distance $D$ between the reactor and the detector increases. This is an
important issue to discriminate a hidden neutron signal from the common
neutron background in the reactor vicinity. Indeed, such a background mainly
depends on local secondary sources.\\

\subsection{Principle of hidden neutron flux detection}

We now want to build a detector that is able to measure the hidden neutron flux.
Neutron detectors are based on the detection of charged particles emitted after 
neutron absorption. In the present case, we suggest the use of a gaseous
detector, such as the usual helium-3 or boron trifluoride (BF$_3$) neutron 
detectors. The mechanism of hidden neutron capture can be described through
the approach of Feinberg and Weinberg \cite{23}, which was also used by Demidov,
Gorbunov, and Tokareva \cite{24} to describe the positronium oscillations in
a volume full of gas in the mirror matter concept. Details of the calculation are
shown in Appendix \ref{GasDet}. We can then compute the event rate $\Gamma $
detected in our brane against the swapping probability $p$
(see Appendix \ref{GasDet}). For monochromatic neutrons, we get the
intuitive but not obvious result

\begin{equation}
\Gamma = \frac 12p\Sigma _A\Phi _{-}V,  \label{rate}
\end{equation}
where $V$ is the volume of the detector and $\Sigma _A$ is the macroscopic
absorption cross section related to BF$_3$ or He-3. For a continuous energy
spectrum, the event rate is obtained by integrating Eq. (\ref{rate}) over
the spectrum. The main difficulty comes from the background which is usually
high in such an environment. The detector must be shielded from neutrons.

\subsection{Proposal at the ILL reactor}

For our experiment, we propose to use the Institute Laue-Langevin (ILL)
reactor (see Fig. 2) where the core (diameter of $40$ cm and height of $80$
cm) is surrounded by a heavy water tank (diameter of $2.5$ m and height of $2
$ m) which will be our hidden neutron source. It can be shown that the major
contribution to the hidden neutron flux comes from the heavy water tank,
while neutrons are strongly absorbed by light water and other surrounding
materials. The thermal neutron flux inside the heavy water tank is modeled
by using a point-like source $\Phi _{+}(r)=S\exp (-r/L)/(4\pi Dr)$, where $%
L=116$ cm and $D=0.57$ cm are the length and coefficient of diffusion 
for heavy water, respectively. $S$ is fitted to match with the known neutron
flux in the heavy water tank \cite{25} with $S=6.0 \times 10^{17}$
neutrons/s. For heavy water, the macroscopic elastic cross section is $%
\Sigma _E=0.4$ cm$^{-1}$. The present rough model is reliable enough to
discuss the experimental concept. Indeed, it is sufficient to assess the
correct magnitude of the induced hidden neutron flux. Of course, any
discussion of the later experimental results will require a detailed
computation of the neutron flux in each location inside of the reactor, as
well as a consideration of the various materials surrounding the core.

We plan to use a cylindrical helium-3 detector (see Fig. 2) with a volume of 
$36$ cm$^3$ and a gas pressure of $4$ atm. This detector could be located
at $10$ meters from the center of the core. The detector will be shielded by
a $4\pi$-box made of a (at least) $3$-cm thick borated rubber ($40\%$ boron content) to capture thermal neutrons, supplemented by a polyethylene
cover with a tunable thickness to moderate epithermal and fast neutrons.
Indeed, hidden neutrons should provide a characteristic constant counting
rate which does not depend on the shielding thickness.

Using Eqs. (\ref{fluxn}) and (\ref{rate}), the estimated rates are shown in Fig. 3
(black solid line). Considering the previous constraint $p<7\times 10^{-6}$ 
\cite{15} (vertical blue solid line), we see that the expected event rate
could be as high as $36$ kHz, which is easily detectable even without
shielding. Actually, we think we can reach an upper rate of about $10$
mHz or even $1$ mHz \cite{26,27,28} with suitable shields and detector
(horizontal red solid line). We can then expect to reach a new upper
constraint for the swapping probability $p$ of about $10^{-\text{9}}$. Such a
constraint would be better than the constraint $p<7\times 10^{-6}$ \cite{15}
by at least $3$ orders of magnitude.

\begin{figure}[h!]
\centerline{\ \includegraphics[width=8.5cm]{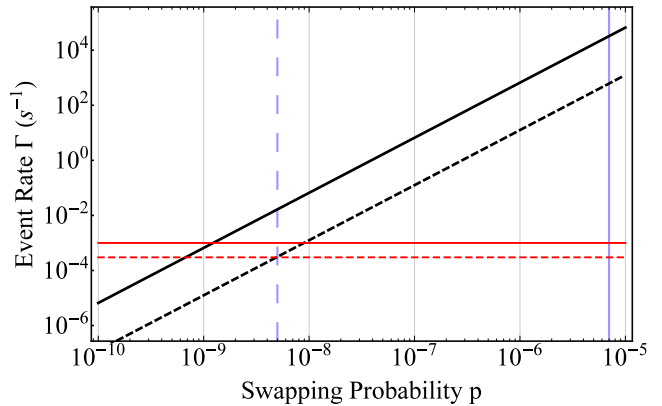}}
\caption{(Color online). Black solid line: Expected event rate $\Gamma$
against the swapping probability $p$. Horizontal red solid line: Expected
threshold of the background noise ($1$ mHz). Vertical blue solid line: Known
upper limit of the constraint on the swapping probability $p <
7\times10^{-6}$ \cite{15}. Black dashed line: Expected event rate $\Gamma$ against
the swapping probability $p$ by considering the experiment in Ref. \cite{26}.
Horizontal red dashed line: Background noise ($0.3$ mHz) in the experiment in Ref. 
\cite{26}. Vertical blue dashed line: Threshold of the
rough constraint on the swapping probability ($p < 5\times10^{-9}$)
considering the experiment in Ref. \cite{26}.}
\label{fig3}
\end{figure}

In Fig. 4, as an example, we show the resulting expected bound on the
coupling constant $g$ between our brane and an invisible one. The values of $%
g$ are given against the gravitational constraint $\eta $ from Eq. (\ref
{Proba}). The existing constraint is given \cite{15} (shaded domain above
the short-dashed blue line) and is compared to the expected results of the 
neutron-shining-through-a-wall experiment (black solid line) by
assuming $A_{amb}\approx 2\times 10^9$\ T$\,$m \cite{15}. For the best rate
constraint, we improve the constraint on $g$ by $2$ orders of magnitude. By
contrast, a constraint on the detected signal lower than $10$ events per
second (see Fig. 3) still allows $p<10^{-7}$, i.e., we improve the constraint
on $g$ by $1$ order of magnitude. As a consequence, the present experiment
can easily improve the constraint on the coupling between our world and
invisible ones.

\begin{figure}[h!]
\centerline{\ \includegraphics[width=8.5cm]{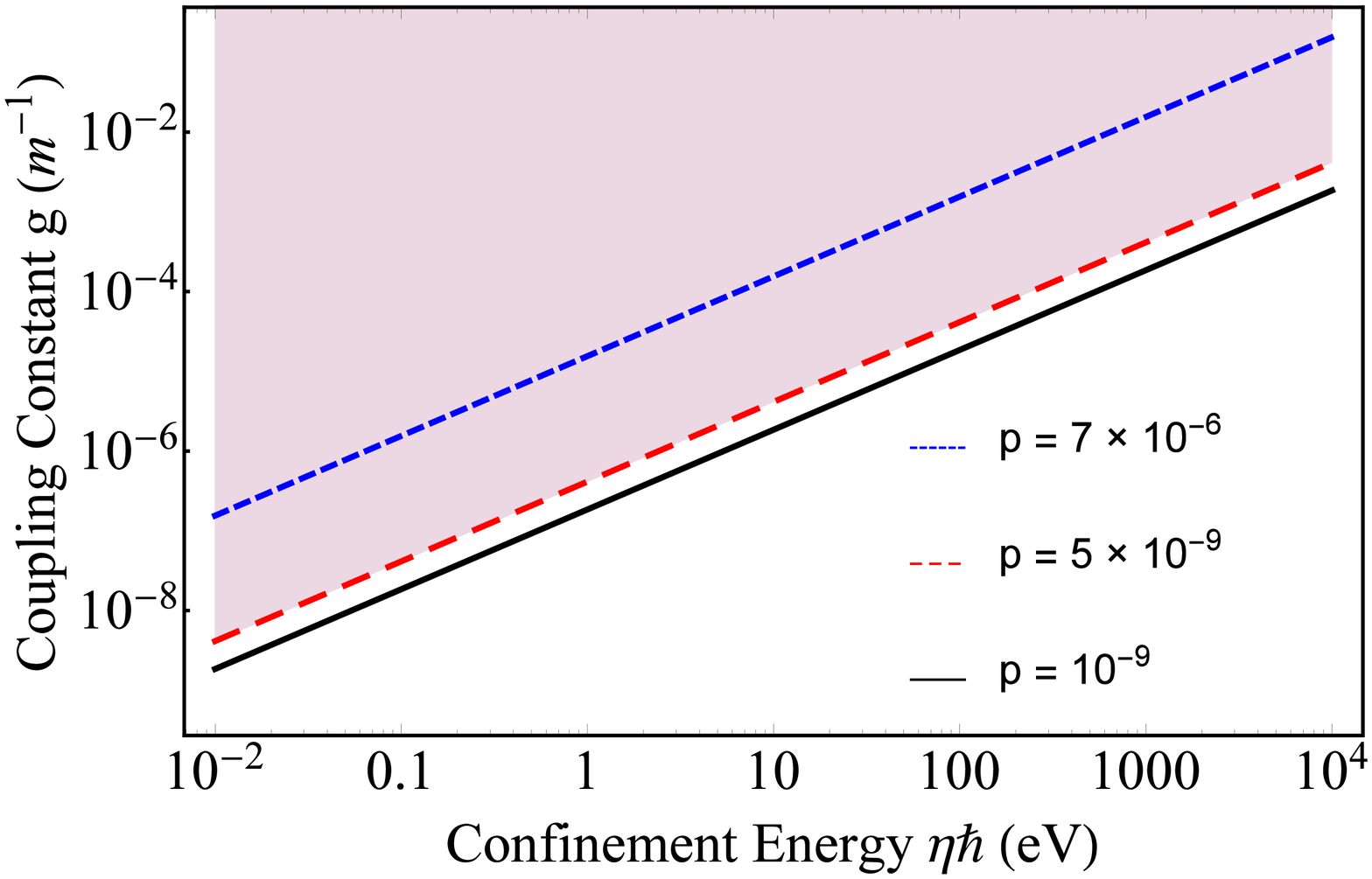}}
\caption{(Color online). Expected experimental limits for the coupling
constant $g$ against the confinement energy for some constraints on the
swapping probability $p$ for $A_{amb}=2\times 10^9$ T m. Red-grey domains
are excluded. $p=7\times 10^{-6}$ corresponds to our previous best
experimental constraint. $p=5\times 10^{-9}$ corresponds to the expected
constraint deduced from the experiment in Ref. \cite{26}.}
\label{fig4}
\end{figure}

\subsection{Yearly time-dependent drift of the swapping probability}

\label{drift}

Let us now briefly underline the possible consequence of the potential $\eta
\hbar $ time evolution  due to the Earth's revolution around the Sun, as
described in section \ref{envpot}. From Eq. (\ref{Proba}) $p$
varies in time as $\Delta p/p=2\Delta \eta /\eta $. Now, from Eqs. (\ref
{fluxn}) and (\ref{rate}) $\Gamma $ varies as $\Delta \Gamma
/\Gamma =2\Delta p/p$. Then, we deduce that the event rate $\Gamma $ varies
in time as $\Delta \Gamma /\Gamma =4\Delta \eta /\eta $, i.e., $\Delta \Gamma
/\Gamma \approx 10^{-3}$ over six months if we consider the values from section 
\ref{envpot}. If it is expected to detect a time-dependent drift of the
swapping probability, for a given duration $\Delta t$ of the experiment, the
number of detected neutrons $N=\Gamma \Delta t$ and its instrumental
uncertainty $\Delta N\sim \sqrt{N}$ imply that enough neutrons must be
detected to allow $\Delta N/N$ to be lower than $4\Delta \eta /\eta $. In
the present case, for month-by-month measurements, we should have $N>10^6$
detected neutrons per month at least. Assuming $\Delta t\sim 20$ days per
month, this leads to $\Gamma \approx 1$ Hz. As a consequence, considering
the above-mentioned detector, a time-dependent drift of the swapping
probability $p$ can be measured provided that $p>3.7\times 10^{-8}$.

\subsection{A constraint as a proof of concept}

A constraint can be suggested from previous experiments \cite{26,27,28}.
Indeed, in different contexts, these experiments required low-noise neutron
detectors to detect ultracold neutrons generated by conventional sources. But
since these detectors were in the nuclear reactor neighborhood, they should
have been sensitive to hidden neutrons as well. As a consequence, considering
the recorded background of these experiments, we can already derive a first
rough constraint on the swapping probability, and thus on the coupling
constant between two adjacent branes. We consider the favorable conditions
introduced in Ref. \cite{26}. The detector volume was $500$ cm$^3$ with a
partial helium-3 pressure of $15$ mbar. The distance between the detector
and the nuclear core was $16.5$ m. The background was $0.3$ mHz \cite{26}. The
resulting constraint is shown in Fig. 3. Using the previously introduced
reactor model, the expected hidden neutron flux $\Gamma$ against the
swapping probability $p$ is shown by the black dashed line. The neutron
background is shown by the horizontal red dashed line. As a result, this
leads to a new constraint for the swapping probability $p$ such that (see
vertical blue dashed line in Fig. 3)

\begin{equation}
p \leq 5\times10^{-9}.  \label{const}
\end{equation}

Such a value can be related to a constraint on the coupling constant between
our visible braneworld and a hypothetical invisible one. This constraint is
given by the red dashed line in Fig. 4. Obviously, these constraints need to
be confirmed or improved by a dedicated experiment such as that described in the
previous sections. As a consequence, the present constraint on $p$ does not
preclude the measurement of a yearly time-dependent drift of the swapping
probability. Nevertheless, it shows the feasibility of the experiment
introduced in the present work.

\section{Conclusion}

We have proposed an experiment in which neutrons can be used to test the
braneworld hypothesis. Using an intense flux of neutrons behind a wall and 
utilizing a properly shielded neutron detector, reappearing neutrons could be
detected as a proof of the existence of hidden braneworlds. Three typical
signatures can be examined to verify the reliability of a measured signal as
a real exotic phenomenon. First, for thick enough shielding, the number of
detected neutrons must not depend on the shield thickness. Second, when the 
distance to the core increases, the number of detected neutrons must decay 
as an inverse-square law against the distance between the nuclear core and
the detector. At last, a yearly modulated time-dependent drift of the number
of detected neutrons could be observed for a not too low swapping
probability. By contrast, without direct detection, such an experiment at
least allows one to constrain the existence of braneworlds. The experimental
constraint on the matter-swapping probability between branes could be
improved by $3$ orders of magnitude with the proposed experiment. Then, a
rough constraint on the swapping probability was proposed to show the
feasibility of a neutron-shining-through-a-wall experiment.

\begin{acknowledgments}
The authors thank Karin Derochette for critical reading of the manuscript.
\end{acknowledgments}

\appendix

\section{Shielded nuclear core as a hidden neutron source}

\label{Scatt}

In a previous work \cite{15}, it was shown that ultracold neutrons
colliding with a solid wall can escape to another hidden brane. In this
situation, since ultracold neutrons present a wavelength roughly larger
than the typical distance between atoms of the wall, the wall appears as a
continuous medium for neutrons \cite{19}. Such a medium can be described
through the Fermi potential \cite{19}. Here, we want to estimate the
production rate of hidden neutrons due to collisions between fast (or
thermal) neutrons and materials surrounding the nuclear core, i.e. mainly
the heavy and light water tanks. In this context, since neutrons cannot be
considered in the ''optical'' domain, the media cannot be considered as
continuous, and we must consider collisions between neutrons and individual
nuclei. Then, our aim is to find a relation between the visible neutron
flux $\Phi _{+}$ and the hidden neutron flux $\Phi _{-}$ induced by the
collisions between visible neutrons and nuclei in our brane. More
specifically, we look for a source term, 
\begin{equation}
S_{-}=K\Phi _{+},  \label{B0}
\end{equation}
where the constant $K$ in the source term will be assessed in the following.
It will be shown that $K=(1/2)p\Sigma _E$, where $\Sigma _E$
is the macroscopic elastic cross section of the materials of the reactor and $\Phi
_{+}$ is the neutron flux in the whole reactor. To do this, considering the
matrix density approach, we follow the Feinberg-Weinberg approach \cite{23}, which was 
also used by Demidov, Gorbunov, and Tokareva \cite{24} to describe other
kinds of oscillations in a volume occupied by matter.

Neutrons are described through the two-brane Pauli equation and we neglect
the neutron decay process: 
\begin{equation}
i\hslash \partial _t\Psi =\mathbf{H}\Psi , \label{B1}
\end{equation}
such that 
\begin{equation}
\mathbf{H}=\left( 
\begin{array}{cc}
-\frac{\hslash ^2}{2m}\Delta +V_{g,+} & -ig\mu _n\mathbf{\sigma \cdot A}%
_{amb} \\ 
ig\mu _n\mathbf{\sigma \cdot A}_{amb} & -\frac{\hslash ^2}{2m}\Delta +V_{g,-}
\end{array}
\right),  \label{B2}
\end{equation}
where $V_{g,+}$ and $V_{g,-}$ are the gravitational potentials felt by the
neutron in each brane. We assume that $\mathbf{\sigma \cdot A}%
_{amb}=A_{amb}\sigma _z$ and we restrain ourselves to a spin-up state
without loss of generality (the matter-swapping probability does not depend
 on the direction of the magnetic vector potential or the spin
state \cite{12,11b,13}). Then we can write
\begin{equation}
\mathbf{H}=\left( 
\begin{array}{cc}
-\frac{\hslash ^2}{2m}\Delta +V_{g,+} & -i\Omega \hbar \\ 
i\Omega \hbar & -\frac{\hslash ^2}{2m}\Delta +V_{g,-}
\end{array}
\right),  \label{B3}
\end{equation}
with $\Omega \hbar =g\mu _nA_{amb}$. The behavior of the neutron flux is
described through the use of the matrix density $\rho$ \cite{23},
\begin{equation}
\frac d{dt}\rho =-i\hbar ^{-1}\left( \mathcal{H}\rho -\rho \mathcal{H}%
^{\dagger }\right) +I_c ,  \label{B4}
\end{equation}
where $I_c$ is the collisional integral which describes the collisions
between neutrons and nuclei in the medium, with 
\begin{equation}
I_c=nv\int F(\theta )\rho F^{\dagger }(\theta )d\Omega .  \label{B5}
\end{equation}
$v$ is the mean relative velocity between neutrons and molecules and $n$
is the density of molecules (nuclei) in matter. We have 
\begin{equation}
F(\theta )=\left( 
\begin{array}{cc}
f(\theta ) & 0 \\ 
0 & 0
\end{array}
\right) , \label{B6}
\end{equation}
which describes the scattering of neutrons by the molecules and where 
$\theta$ is a scattering angle. The second diagonal term is equal to zero
since we assume there is no matter in the hidden brane.

We also define $\mathcal{H=}\mathbf{H}\mathcal{+}\mathbf{C}$, with 
\begin{equation}
\mathbf{C}=\left( 
\begin{array}{cc}
-2\pi nv\hslash f(0)/k & 0 \\ 
0 & 0
\end{array}
\right),  \label{B7}
\end{equation}
(where $k$ is the neutron wave vector), which accounts for the presence of matter in the system. Let us use the definition 
\begin{equation}
\rho =\left( 
\begin{array}{cc}
\rho _{+} & r+is \\ 
r-is & \rho _{-}
\end{array}
\right) .  \label{B8}
\end{equation}
We also define 
\begin{equation}
w_R=4\pi \frac{nv}k\text{Re}(f(0)),\text{ and }w_I=4\pi \frac{nv}k\text{Im}%
(f(0))  \label{B9}
\end{equation}
with the optical theorem related to the cross section $\sigma _{tot}$: 
\begin{equation}
\sigma _{tot}=\sigma _E+\sigma _I=\frac{4\pi }k\text{Im}(f(0))=\frac{w_I}{nv},
\label{B10}
\end{equation}
where $\sigma _E$ and $\sigma _I$ are the elastic and inelastic (absorption) cross sections, and 
\begin{equation}
\sigma _E=\int \left| f(\theta )\right| ^2d\Omega.  \label{B11}
\end{equation}
We then obtain the following system from Eq. (\ref{B4}): 
\begin{eqnarray}
\frac d{dt}\rho _{+}=-2\Omega r-nv\sigma _I\rho _{+},  \label{B12}
\end{eqnarray}
\begin{equation}
\frac d{dt}\rho _{-}=2\Omega r,  \label{B13}
\end{equation}
\begin{equation}
\frac d{dt}r=\left( \eta -(1/2)w_R\right) s-(1/2)rw_I+\Omega \left( \rho
_{+}-\rho _{-}\right) ,  \label{B14}
\end{equation}
\begin{equation}
\frac d{dt}s=-\left( \eta -(1/2)w_R\right) r-(1/2)sw_I,  \label{B15}
\end{equation}
with $\hbar \eta =V_{g,+}-V_{g,-}$. In the following, we consider a
statistical set of neutrons initially localized in our visible brane and
emerging from the nuclear core, i.e., $\rho _{+}(t=0)=1$ and $\rho _{-}(t=0)=0
$. If we consider Eq. (\ref{B12}), we obviously deduce that the neutron
population in our brane must mainly decrease due to absorption, i.e., neutron
leakage into a hidden brane must be very weak to be consistent with known
physics. So, we must verify that $2\Omega r\ll nv\sigma _I\rho _{+}$. Now, if (for instance) we
consider thermal neutrons in heavy water, and if we consider $%
\Omega $ with $A_{amb}=2\times 10^9$ T m and $g=10^{-3}$ m$^{-1}$, we get $%
nv\sigma _I\approx 7.3$ s$^{-1}$ and $\Omega \approx 2\times 10^{14}$ s$^{-1}
$, i.e., 2$\Omega \gg nv\sigma _I$. As a consequence, the previous condition
must imply that $r\ll \rho _{+}$. Since neutron leakage must be weak, we also
deduce that $\rho _{-}\ll \rho _{+}$. At last, assuming that $\eta \gg w_R$ and
using Eq. (\ref{B10}), Eqs. (\ref{B12})-(\ref{B15}) can be recast in the
following convenient form: 
\begin{eqnarray}
\frac d{dt}\rho _{+}=-nv\sigma _I\rho _{+},  \label{B12b}
\end{eqnarray}
\begin{equation}
\frac d{dt}\rho _{-}=2\Omega r,  \label{B13b}
\end{equation}
\begin{equation}
\frac d{dt}r=\eta s-(1/2)nv\sigma _{tot}r+\Omega \rho _{+},  \label{B14b}
\end{equation}
\begin{equation}
\frac d{dt}s=-\eta r-(1/2)nv\sigma _{tot}s.  \label{B15b}
\end{equation}

At this stage of the problem, it is relevant to underline that from Eqs. (%
\ref{B14b}) and (\ref{B15b}) we can easily deduce
\begin{equation}
\frac d{dt}m=-(1/2)nv\sigma _{tot}m+\Omega \cos \theta \rho _{+} , \label{B16}
\end{equation}
where we have set $r+is=me^{i\theta }$. Without coupling, i.e., if $\Omega
=0 $, we deduce from Eq. (\ref{B16}) that the coherence terms of the density
matrix (i.e., its off-diagonal terms $r$ and $s$) must exponentially decay as 
$\exp (-(1/2)nv\sigma _{tot}t)$. As a consequence, the collisional dynamics
should suppress the quantum coherence precluding the neutron swapping
between branes. By contrast, Eq. (\ref{B16}) as well as Eqs. (\ref{B14b})
and (\ref{B15b}) also show us that the coupling $\Omega$ can prevent 
decoherence. This allows stationary coherences such that $dr/dt=ds/dt=0$,
with $r$ and $s$ different from zero. The steady state is quickly achieved
due to the above-mentioned exponential behavior. Let us call $X$ the
distance covered by the neutron in the medium. The length $X$ cannot be
equated to the crossed thickness $L$ in the medium. Since neutrons are
localized in our visible brane, they can be diffused by molecules or 
the nuclei of atoms. $X$ is the sum of the lengths of the various straight lines covered
by the neutron, and thus $L=X\sqrt{\sigma _I/3\sigma _E}$. Now, for instance,
since $n\sigma _{tot}\approx 0.4$ cm$^{-1}$ for heavy water, the stationary
coherences occur at $X=4$ cm, i.e., $L=0.2$ mm from the neutron source, a
distance which must be compared to the $105$-cm thick heavy water slab in
the ILL's reactor. By contrast, $n\sigma _I=4\times 10^{-5}$ cm$^{-1}$ in D$%
_2$O, i.e., $\rho _{+}$ slowly varies in the heavy water region (see Eq. (\ref
{B12b})). Then, in the stationary coherences hypothesis, Eqs. (\ref{B14b})
and (\ref{B15b}) become a linear system of two equations with two unknowns ($%
r$ and $s$) which is trivially solved. We get

\begin{equation}
\left( 
\begin{array}{c}
r \\ 
s
\end{array}
\right) =\frac 1{(1/4)\left( nv\sigma _{tot}\right) ^2+\eta ^2}\left( 
\begin{array}{c}
(1/2)nv\sigma _{tot}\Omega \rho _{+} \\ 
-\eta \Omega \rho _{+}
\end{array}
\right).  \label{B17}
\end{equation}
From Eqs. (\ref{B13b}) and (\ref{B17}) we deduce 
\begin{equation}
\frac d{dt}\rho _{-}\approx (1/2)pnv\sigma _{tot}\rho _{+},  \label{B18}
\end{equation}
where we have set $p=2\Omega ^2/\eta ^2$ to be consistent with
 Eq. (\ref{Proba}). Let us now consider an initial local neutron flux $\Phi _0=u_0v$ where $%
u_0$ is the initial neutron density. The neutron fluxes in each brane are
given by $\Phi _{\pm }=\Phi _0\rho _{\pm }$. Then, from Eq. (\ref
{B18}) we deduce 
\begin{equation}
\partial _t\Phi _{-}=(1/2)pnv\sigma _{tot}\Phi _{+}.  \label{B18bis}
\end{equation}
Since the neutron flux $\Phi _{-}=u_{-}v$, where $u_{-}$ is the local
neutron density in the hidden brane, then
\begin{equation}
\partial _tu_{-}=(1/2)pn\sigma _{tot}\Phi _{+}.  \label{B19}
\end{equation}
Since $u_{-}$ is now local, Eq. (\ref{B19}) must be supplemented by a
divergence term $\mathbf{\nabla \cdot j}_{-}$ to account for the local
behavior of the neutron current $\mathbf{j}_{-}=u_{-}\mathbf{v}$. We then
deduce the continuity equation for neutrons in the second brane,

\begin{equation}
\mathbf{\nabla \cdot j}_{-}+\partial _tu_{-}=(1/2)pn\sigma _{tot}\Phi _{+},
\label{B20}
\end{equation}
which is the continuity equation endowed with a source term $S_{-}$, 
\begin{equation}
S_{-}=(1/2)pn\sigma _{tot}\Phi _{+} . \label{B21}
\end{equation}
Obviously, for heavy water, $\sigma _{tot}\approx \sigma _E$ and we retrieve
Eq. (\ref{B0}) since $n\sigma _E = \Sigma_E$. Now, as a striking result, we note that neutrons propagating
in the hidden brane are not absorbed since we expect vacuum rules in
the other world. As a consequence, while visible neutrons have been stopped
by the water surrounding the nuclear core, hidden neutrons can propagate
freely far away from the reactor. Then we can completely suppress the
visible neutron flux in our brane while keeping a constant flux of hidden
neutrons in the invisible brane. A hidden neutron source from a nuclear core
surrounded by a relevant shield can then be built.

\section{Hidden neutron detector}

\label{GasDet}

We consider a gas stored in a vessel as a neutron detector. More
specifically, we consider a neutron flux in the second brane and we expect
to detect neutrons which arise from the second brane in our braneworld. The
distance between the atoms in the gas is much larger than the neutrons' 
wavelength. As a consequence, neutrons see a set of scattering objects
instead of a continuous medium. We follow the same approach as before,
and start with Eqs. (\ref{B12}) to (\ref{B15}), but now we consider a
statistical set of neutrons initially localized in the invisible brane, such
as it can be written $\rho _{+}(t=0)=0$ and $\rho _{-}(t=0)=1$.

Looking at Eq. (\ref{B13}), we see that the neutron population in the hidden
brane mainly decreases due to neutron leakage into a hidden brane. Following
a similar hypothesis as in the previous section, we can consider that $%
2\Omega r$ must be very weak and we assume that $\partial _t\rho _{-}\approx
0$, i.e., $\rho _{-}$\ is almost constant at the detector scale. We can also
assume that $\rho _{+}\ll \rho _{-}$ and $\eta \gg w_R$. Then (using Eq. (\ref
{B10})), Eqs. (\ref{B12})-(\ref{B15}) can be recast in the following
convenient form:

\begin{eqnarray}
\frac d{dt}\rho _{+}=-2\Omega r-nv\sigma _I\rho _{+},  \label{C1}
\end{eqnarray}
\begin{equation}
\rho _{-}\approx 1,  \label{C2}
\end{equation}
\begin{equation}
\frac d{dt}r=\eta s-(1/2)rnv\sigma _{tot}-\Omega \rho _{-},  \label{C3}
\end{equation}
\begin{equation}
\frac d{dt}s=-\eta r-(1/2)snv\sigma _{tot}.  \label{C4}
\end{equation}

Obviously, $\sigma _{tot}$, $\sigma _I$, and $\sigma _E$ are now the cross sections 
related to the gas and $n$ is the density of gas molecules.

Now, from Eqs. (\ref{C3}) and (\ref{C4}) we can easily deduce that 
\begin{equation}
\frac d{dt}m=-(1/2)nv\sigma _{tot}m-\Omega \cos \theta \rho _{-} , \label{C5}
\end{equation}
where $r+is=me^{i\theta }$ is set. Without coupling, i.e., if $\Omega =0 $,
we deduce from Eq. (\ref{C5}) that the coherence terms of the density matrix
(i.e., its off-diagonal terms $r$ and $s$) must exponentially decay as $\exp
(-(1/2)nv\sigma _{tot}t)$. As a consequence, the collisional dynamics should
suppress the quantum coherence precluding the neutron swapping between
branes. By contrast, Eq. (\ref{C5}) as well as Eqs. (\ref{C3}) and (\ref{C4}%
) also show that the coupling $\Omega $ preserves from decoherence.
Nevertheless, in contrast to the case described in the previous section,
stationary coherences such that $dr/dt=ds/dt=0$ (with $r$ and $s$ different
from zero) cannot be achieved here due to the short length of the detector.
Indeed, assuming for instance a He-3 detector with a pressure of about $4$ atm, 
we get $n\sigma _{tot}\approx 0.53$ cm$^{-1}$ for thermal neutrons. Now, the
distance $X=vt$ crossed by a neutron can be equated to the length $L$ of
the detector since we can assume that the hidden neutrons are not
significantly diffused by the nuclei of gas molecules or atoms in our brane.
Let us consider (for instance) a detector length of about $10$ cm. At half length, we
get $\exp (-(1/2)nv\sigma _{tot}t)\approx 27$ $\%$, i.e., the coherences are
not negligible in at least half of the detector. As a consequence,
we cannot apply the stationary coherences hypothesis here.

Nevertheless, Eqs.(\ref{C3}) and (\ref{C4}) constitute a simple
nonhomogeneous first-order differential equation that is easy to solve since $\rho
_{-}\approx 1$, i.e., since $\rho _{-}$ is constant. Let us set

%\begin{widetext}
\begin{equation}
\left( 
\begin{array}{c}
r \\ 
s
\end{array}
\right) =e^{-(1/2)nv\sigma _{tot}t}\left( 
\begin{array}{cc}
\cos (\eta t) & \sin (\eta t) \\ 
-\sin (\eta t) & \cos (\eta t)
\end{array}
\right) \left( 
\begin{array}{c}
R \\ 
S
\end{array}
\right) . \label{C6}
\end{equation}
If we insert Eq. (\ref{C6}) into Eqs.(\ref{C3}) and (\ref{C4}), we obtain 
\begin{equation}
\frac d{dt}\left( 
\begin{array}{c}
R \\ 
S
\end{array}
\right) =e^{(1/2)nv\sigma _{tot}t}\left( 
\begin{array}{cc}
\cos (\eta t) & -\sin (\eta t) \\ 
\sin (\eta t) & \cos (\eta t)
\end{array}
\right) \left( 
\begin{array}{c}
-\Omega \rho _{-} \\ 
0
\end{array}
\right) , \label{C7}
\end{equation}

from which we deduce that 
\begin{equation}
\left( 
\begin{array}{c}
R \\ 
S
\end{array}
\right) =-\Omega \rho _{-}\left( 
\begin{array}{c}
\int_0^t\cos (\eta t^{\prime })e^{(1/2)nv\sigma _{tot}t^{\prime }}dt^{\prime
} \\ 
\int_0^t\sin (\eta t^{\prime })e^{(1/2)nv\sigma _{tot}t^{\prime }}dt^{\prime
}
\end{array}
\right).  \label{C8}
\end{equation}
%\end{widetext}

Using Eqs.(\ref{C6}) and (\ref{C8}), Eq. (\ref{C1}) can be rewritten as 
\begin{eqnarray}
\frac d{dt}\rho _{+} &=&2\Omega ^2\rho _{-}\int_0^te^{-(1/2)nv\sigma
_{tot}(t-t^{\prime })}\cos (\eta (t-t^{\prime }))dt^{\prime }  \nonumber \\
&&-nv\sigma _I\rho _{+},  \label{C9}
\end{eqnarray}
which leads to

\begin{widetext}
\begin{equation}
\rho _{+}=2\Omega ^2\rho _{-}\int_0^t\int_0^{t^{\prime }}e^{-(1/2)nv\sigma
_{tot}(t^{\prime }-t^{\prime \prime })}e^{-nv\sigma _I(t-t^{\prime })}\cos
(\eta (t^{\prime }-t^{\prime \prime }))dt^{\prime \prime }dt^{\prime }.
\label{C10}
\end{equation}
\end{widetext}

From the properties of the density matrix $\rho $, we can deduce the
emerging neutron flux $\Phi _v$ which can be detected from the hidden
current: $\Phi _v=\Phi _{-}\rho _{+}$. Considering He-$3$ mixtures or BF$_3$%
, we assume that the inelastic cross section $\sigma _I$ stands for the
neutron capture probability and the production of protons, which can be
recorded. Then, the rate $\Gamma $ of events recorded per second by the
detector in our brane is given by
\begin{eqnarray}
\Gamma &=&\int_0^{L/v}nv\sigma _I\;\int_{S_d}\Phi _vdSdt  \label{C12} \\
&=&\int_0^{L/v}nv\sigma _I\rho _{+}\;\int_{S_d}\Phi _{-}dSdt , \nonumber
\end{eqnarray}
where $S_d$ is the effective area of the detector and $L$ is the length of
the gas vessel, such that $L/v$ is the time during which hidden neutrons
travel in the detector volume. The length $L$ can be equated to the
length of the detector since we can assume that the hidden neutrons are not
significantly diffused by the nuclei of gas molecules or atoms. In addition,
absorption prevails on diffusion processes ($\sigma _I\gg \sigma _E$ in
helium-3 or BF$_3$).

\begin{widetext}
Assuming that $\eta \gg nv\sigma _{tot}$, and setting $p=2\Omega ^2/\eta ^2$ and $x=vt$, from Eqs. (\ref{C10}) and (\ref{C12}) we get
\begin{equation}
\Gamma =\varphi _sp\frac{n\sigma _I\eta ^2}{v^2}\int_0^Le^{-n\sigma
_Ix}\int_0^x\int_0^{x^{\prime }}e^{(1/2)n\sigma _{tot}(x^{\prime \prime
}-x^{\prime })}e^{n\sigma _Ix^{\prime }}\cos \left( (\eta /v)\left(
x^{\prime \prime }-x^{\prime }\right) \right) dx^{\prime \prime }dx^{\prime
}dx  \label{C13}
\end{equation}

with $\varphi _s=\int_{S_d}\Phi _{-}dS$.
\end{widetext}

Due to the fast oscillating cosine term, and since $\eta \gg nv\sigma _{tot}$%
, Eq. (\ref{C13}) reduces to
\begin{equation}
\Gamma \sim (1/2)\varphi _sp\left[ (\frac{\sigma _E}{\sigma _I}%
-1)(e^{-n\sigma _IL}-1)+n\sigma _{tot}L\right].  \label{C14}
\end{equation}
For pure helium-3, relevant argon-helium-3 mixtures, or BF$_3$, we verify that 
$\sigma _I\gg \sigma _E$, and then for long enough detectors we simply get the
intuitive but not so obvious result 
\begin{equation}
\Gamma \sim (1/2)pn\sigma _I V\Phi _{-} , \label{C15}
\end{equation}
where $V=S_d L$ is the volume of the detector and $n\sigma _I = \Sigma _A$ is the
macroscopic absorption cross section of the gas.


\begin{thebibliography}{99}
\bibitem{1}  K. Akama, \textit{Pregeometry}, Lect. Notes Phys. \textbf{176}
(1983) 267, arXiv:hep-th/0001113; \\V.A. Rubakov, M.E. Shaposhnikov, \textit{%
Do we live inside a domain wall?}, Phys. Lett. \textbf{125B} (1983) 136;\\M.
Pavsic, \textit{Einstein's gravity from a first order lagrangian in an
embedding space}, Phys. Lett. \textbf{116A} (1986) 1, arXiv:gr-qc/0101075; \\%
P. Horava, E. Witten, \textit{Heterotic and Type I String Dynamics from
Eleven Dimensions}, Nucl. Phys. \textbf{B460} (1996) 506 ,
arXiv:hep-th/9510209;\\A. Lukas, B.A. Ovrut, K.S. Stelle, D. Waldram, 
\textit{The Universe as a Domain Wall}, Phys. Rev. D \textbf{59} (1999)
086001, arXiv:hep-th/9803235; \\R. Davies, D. P. George, R. R. Volkas, 
\textit{The standard model on a domain-wall brane?}, Phys. Rev. D \textbf{77}
(2008) 124038, arXiv:0705.1584 [hep-ph].

\bibitem{2}  P. Brax, C. van de Bruck, A.-C. Davis, \textit{Brane world
cosmology}, Rep. Prog. Phys. \textbf{67} (2004) 2183-2231,
arXiv:hep-th/0404011;\\P. Brax, C. van de Bruck, \textit{Cosmology and brane
worlds: a review}, Class. Quantum Grav. \textbf{20} (2003) R201-R232,
arXiv:hep-th/0303095;\\R. Dick, \textit{Brane worlds}, Class. Quantum Grav. 
\textbf{18} (2001) R1-R23, arXiv:hep-th/0105320.

\bibitem{3}  I. Antoniadis, N. Arkani-Hamed, S. Dimopoulos, G. Dvali, 
\textit{New dimensions at a millimeter to a Fermi and superstrings at a TeV}%
, Phys. Lett. B \textbf{436} (1998) 257, arXiv:hep-ph/9804398;\\N.
Arkani-Hamed, S. Dimopoulos, G. Dvali, \textit{Phenomenology, Astrophysics
and Cosmology of Theories with Sub-Millimeter Dimensions and TeV Scale
Quantum Gravity}, Phys. Rev. D \textbf{59} (1999) 086004,
arXiv:hep-ph/9807344.

\bibitem{4}  D. Hooper, S. Profumo, \textit{Dark Matter and Collider
Phenomenology of Universal Extra Dimensions}, Phys. Rep. \textbf{453} (2007)
29, arXiv:hep-ph/0701197.

\bibitem{5}  J.A.R. Cembranos, R. L. Delgado, A. Dobado, \textit{Brane
worlds at the LHC: Branons and KK gravitons}, Phys. Rev. D \textbf{88},
075021 (2013).

\bibitem{6}  I. Antoniadis, S. Baessler, M. B\"{u}chner, V.V. Fedorov, S.
Hoedl, A. Lambrecht, V.V. Nesvizhevsky, G. Pignol, K.V. Protasov, S.
Reynaud, Yu. Sobolev, \textit{Short-range fundamental forces}, C. R.
Physique \textbf{12} (2011) 755-778.

\bibitem{7}  J. Redondo, A. Ringwald, \textit{Light shining through walls},
Contemp. Phys. \textbf{52} (2011) 211-236, arXiv:1011.3741 [hep-ph].

\bibitem{7b}  S.N. Gninenko, \textit{Search for MeV dark photons in a
light-shining-through-walls experiment at CERN}, Phys. Rev. D \textbf{89}
(2014) 075008, arXiv:1308.6521 [hep-ph].

\bibitem{8a}  S.N. Gninenko, N.V. Krasnikov, V.A. Matveev, \textit{Invisible
decay of muonium: Tests of the standard model and searches for new physics},
Phys. Rev. D \textbf{87} (2013) 015016, arXiv:1209.0060 [hep-ph].

\bibitem{8b} S.N. Gninenko, \textit{Search for invisible decays of $\pi^0, \eta, \eta', K_S$ and $K_L$: A probe of new physics and 
tests using the Bell-Steinberger relation}, Phys. Rev. D \textbf{91} (2015) 015004, arXiv:1409.2288 [hep-ph].

\bibitem{8}  H. Abele, \textit{The neutron. Its properties and basic
interactions}. Prog. Part. Nucl. Phys. \textbf{60} (2008) 1.

\bibitem{11}  M. Sarrazin, F. Petit, \textit{Equivalence between domain
walls and ''noncommutative'' two-sheeted spacetimes: Model-independent
matter swapping between branes}, Phys. Rev. D \textbf{81}, 035014 (2010),
arXiv:0903.2498 [hep-th].

\bibitem{12}  F. Petit, M. Sarrazin, \textit{Quantum dynamics of massive
particles in a non-commutative two-sheeted space-time}, Phys. Lett. B 
\textbf{612}, 105 (2005), arXiv:hep-th/0409084.

\bibitem{11b}  M. Sarrazin, F. Petit, \textit{Brane matter, hidden or mirror
matter, their various avatars and mixings: many faces of the same physics},
Eur. Phys. J. C \textbf{72} (2012) 2230, arXiv:1208.2014 [hep-ph].

\bibitem{13}  M. Sarrazin, F. Petit, \textit{Matter localization and
resonant deconfinement in a two-sheeted spacetime}, Int. J. Mod. Phys. A 
\textbf{22}, 2629 (2007), arXiv:hep-th/0603194.

\bibitem{14}  M. Sarrazin, F. Petit, \textit{Laser frequency combs and
ultracold neutrons to probe braneworlds through induced matter swapping
between branes}, Phys. Rev. D \textbf{83}, 035009 (2011), arXiv:0809.2060
[hep-ph].

\bibitem{15}  M. Sarrazin, G. Pignol, F. Petit, V.V. Nesvizhevsky, \textit{%
Experimental limits on neutron disappearance into another braneworld}, Phys.
Lett. B \textbf{712} (2012) 213, arXiv:1201.3949 [hep-ph].

\bibitem{15b}  S. Baessler, V.V. Nesvizhevsky, K.V. Protasov, A.Yu. Voronin, 
\textit{Constraint on the coupling of axionlike particles to matter via an
ultracold neutron gravitational experiment}, Phys. Rev. D \textbf{75},
075006 (2007), arXiv:hep-ph/0610339.

\bibitem{15c}  I. Altarev, C.A. Baker, G. Ban, K. Bodek, M. Daum, P.
Fierlinger, P. Geltenbort, K. Green, M.G.D. van der Grinten, E. Gutsmiedl,
P.G. Harris, R. Henneck, M. Horras, P. Iaydjiev, S. Ivanov, N. Khomutov, K.
Kirch, S. Kistryn, A. Knecht, P. Knowles, A. Kozela, F. Kuchler, M.
Kuz'niak, T. Lauer, B. Lauss, T. Lefort, A. Mtchedlishvili, O.
Naviliat-Cuncic, S. Paul, A. Pazgalev, J.M. Pendlebury, G. Petzoldt, E.
Pierre, C. Plonka-Spehr, G. Qu\'{e}m\'{e}ner, D. Rebreyend, S. Roccia, G.
Rogel, N. Severijns, D. Shiers, Yu. Sobolev, R. Stoepler, A. Weis, J. Zejma,
J. Zenner, G. Zsigmond,\textit{Neutron to Mirror-Neutron Oscillations in the
Presence of Mirror Magnetic Fields}, Phys. Rev. D \textbf{80}, 032003
(2009), arXiv:0905.4208 [nucl-ex];\\A.P. Serebrov, E.B. Aleksandrov, N.A.
Dovator, S.P. Dmitriev, A.K. Fomin, P. Geltenbort, A.G. Kharitonov, I.A.
Krasnoschekova, M.S. Lasakov, A.N. Murashkin, G.E. Shmelev, V.E. Varlamov,
A.V. Vassiljev, O.M. Zherebtsov, O. Zimmer, \textit{Experimental search for
neutron - mirror neutron oscillations using storage of ultracold neutrons},
Phys. Lett. B \textbf{663}, 181 (2008), arXiv:0706.3600 [nucl-ex];\\G. Ban,
K. Bodek, M. Daum, R. Henneck, S. Heule, M. Kasprzak, N. Khomutov, K. Kirch,
S. Kistryn, A. Knecht, P. Knowles, M. Kuzniak, T. Lefort, A. Mtchedlishvili,
O. Naviliat-Cuncic, C. Plonka, G. Qu\'{e}m\'{e}ner, M. Rebetez, D.
Rebreyend, S. Roccia, G. Rogel, M. Tur, A. Weis, J. Zejma, G. Zsigmond , 
\textit{Direct Experimental Limit on Neutron Mirror-Neutron Oscillations},
Phys. Rev. Lett. \textbf{99}, 161603 (2007), arXiv:0705.2336 [nucl-ex].

\bibitem{15d}  Y.N. Pokotilovski, \textit{On the experimental search for
neutron - mirror neutron oscillations}, Phys.Lett. B \textbf{639} (2006)
214-217, arXiv:nucl-ex/0601017.

\bibitem{15e}  Z. Berezhiani, \textit{More about neutron - mirror neutron
oscillation}, Eur. Phys. J. C \textbf{64} (2009) 421, arXiv:0804.2088 [hep-ph]; \\Z.
Berezhiani, L. Bento, \textit{Neutron-Mirror Neutron Oscillations: How Fast
Might They Be?}, Phys. Rev. Lett. \textbf{96} (2006) 081801, arXiv:hep-ph/0507031.

\bibitem{15ee} R.N. Mohapatra, S. Nasri, S. Nussinov, \textit{Some Implications of Neutron Mirror Neutron Oscillation}, 
Phys. Lett. B \textbf{627} (2005) 124, arXiv:hep-ph/0508109.

\bibitem{15f}  G. Dvali, M. Redi, \textit{Phenomenology of 10$^{32}$ Dark
Sectors}, Phys. Rev. D \textbf{80} (2009) 055001, arXiv:0905.1709 [hep-ph].

\bibitem{9}  A.T. Yue, M.S. Dewey, D.M. Gilliam, G.L. Greene, A.B. Laptev,
J.S. Nico, W.M. Snow, F.E. Wietfeldt, \textit{Improved Determination of the
Neutron Lifetime}, Phys. Rev. Lett. \textbf{111}, 222501 (2013),
arXiv:1309.2623 [nucl-ex].

\bibitem{10}  A. Coc, M. Pospelov, J.-P. Uzan, E. Vangioni, \textit{Modified
big bang nucleosynthesis with non-standard neutron sources}, Phys. Rev. D 
\textbf{90}, 085018 (2014), arXiv:1405.1718 [hep-ph].

\bibitem{16}  R. Lakes, \textit{Experimental Limits on the Photon Mass and
Cosmic Magnetic Vector Potential}, Phys. Rev. Lett. \textbf{80}, 1826 (1998);%
\\J. Luo, C.-G. Shao, Z.-Z. Liu, Z.-K. Hu, \textit{Determination of the
limit of photon mass and cosmic magnetic vector with rotating torsion balance%
}, Phys. Lett. A \textbf{270}, 288 (2000).

\bibitem{23}  G. Feinberg, S. Weinberg, \textit{Conversion of muonium into
antimuonium}, Phys. Rev. \textbf{123}, 1439 (1961).

\bibitem{24}  S.V. Demidov, D.S. Gorbunov, A.A. Tokareva, \textit{%
Positronium oscillations to Mirror World revisited}, Phys. Rev. D \textbf{85}%
, 015022 (2012), arXiv:1111.1072 [hep-ph].

\bibitem{25}  D. Ridikas, G. Fioni, P. Goberis, O. Deruelle, M. Fadil, F.
Marie, S. R\"{o}ttger, \textit{On the fuel cycle and neutron fluxes of the
high flux reactor at ILL Grenoble}, Proc. of the 5th Int. Specialists'
meeting SATIF-5, OECD/NEA Paris (2000).

\bibitem{26}  V.V. Nesvizhevsky, E.V. Lychagin, A.Yu. Muzychka, G.V.
Nekhaev, A.V. Strelkov, \textit{About interpretation of experiments on small
increase in energy of UCN in traps}, Physics Letters B \textbf{479} (2000)
353.

\bibitem{27}  E.V. Lychagin, A.Yu. Muzychka, V.V. Nesvizhevsky, G.V.
Nekhaev, A.V. Strelkov, \textit{Experimental estimation of the possible
subbarrier penetration of ultracold neutrons through vacuum-tight foils},
JETP Letters \textbf{71} (2000) 447.

\bibitem{28}  V.V. Nesvizhevsky, H.G. Boerner, A.M. Gagarski, A.K.
Petoukhov, G.A. Petrov, H. Abele, S. Baessler, G. Divkovic, F.J. Ruess, T.
Stoeferle, A. Westphal, A.V. Strelkov, K.V. Protasov, A.Yu. Voronin, \textit{%
Measurement of quantum states of neutrons in the Earth's gravitational field}%
, Phys. Rev. D \textbf{67} (2003) 102002, arXiv:hep-ph/0306198.

\bibitem{19}  A.-J. Dianoux, G. Lander, \textit{Neutron Data Booklet}, (OCP
Science, 2003).
\end{thebibliography}
\end{document}